\documentclass[12pt]{article}

\newcommand{\VersionInformation}{}  
\InputIfFileExists{Debug}{}{}

\usepackage[utf8]{inputenc}
\usepackage{amsmath}
\usepackage{amsthm}
\usepackage{amsfonts}          
\usepackage{amssymb}           
\usepackage[mathscr]{eucal}
\usepackage{graphicx}
\usepackage{color}
\usepackage[normalem]{ulem}
\usepackage{hypbmsec}
\usepackage{fancyvrb}
\usepackage{sepnum}
\usepackage{xspace}

\usepackage{rotating}
\usepackage[isu,bf]{caption}
\usepackage{multirow}

\newlength{\xtrawidth}
\newlength{\xtraheight}
\usepackage[all]{xy}           

\usepackage[backref,linktocpage,bookmarks]{hyperref}

\ifx\DEBUG\EMPTYMACRO
\else
  \usepackage{showlabels}
\fi

\def\clap#1{\hbox to 0pt{\hss#1\hss}}

\def\mathclap{\mathpalette\mathclapinternal}

\def\mathclapinternal#1#2{%
\clap{$\mathsurround=0pt#1{#2}$}}	

\makeatletter
  \def\adots{\mathinner{\mkern2mu\raise\p@\hbox{.}
      \mkern2mu\raise4\p@\hbox{.}\mkern1mu
      \raise7\p@\vbox{\kern7\p@\hbox{.}}\mkern1mu}}
\makeatother

\newcommand{\eqdef}{%
  \mathrel{\lower.1mm
    \hbox{$\stackrel{\lower.424ex\hbox{\scriptsize def}}{=}$}}
}
\newcommand{\Q}{\ensuremath{{\mathbb{Q}}}}

\newcommand{\C}{\ensuremath{{\mathbb{C}}}}

\newcommand{\Z}{\mathbb{Z}}
\newcommand{\CP}{{\ensuremath{\mathop{\null {\mathbb{P}}}\nolimits}}}

\DeclareMathOperator{\Span}{span}

\DeclareMathOperator{\Aut}{Aut}

\DeclareMathOperator{\Hom}{Hom}

\DeclareMathOperator{\Sing}{Sing}

\DeclareMathOperator{\diag}{diag}

\newcommand{\Xt}{{\ensuremath{\widetilde{X}}}}

\newcommand{\Osheaf}{\ensuremath{\mathscr{O}}}

\newcommand{\dual}{\ensuremath{\vee}}

{\end{list}}

{\everymath{\displaystyle\everymath{}}\array}%
{\endarray}

\pdfoutput=1

\DeclareMathOperator{\conv}{conv}
\DeclareMathOperator{\Spec}{\mathop{\mathrm{Spec}}}

\newcommand{\emptycone}{{\ensuremath{\langle\rangle}}}
\newcommand{\fan}{\ensuremath{\mathcal{F}_\nabla}}

\begin{document}
\begin{titlepage}
  \vspace*{-2cm}
  \VersionInformation
  \hfill
  \parbox[c]{5cm}{
    \begin{flushright}
    \end{flushright}
  }
  \vspace*{2cm}
  \begin{center}
    \Huge 
    The 24-Cell and Calabi-Yau Threefolds with Hodge Numbers (1,1)
  \end{center}
  \vspace*{8mm}
  \begin{center}
    \begin{minipage}{\textwidth}
      \begin{center}
        \sc 
        Volker Braun
      \end{center}
      \begin{center}
        \textit{
          Dublin Institute for Advanced Studies\hphantom{${}^1$}\\
          10 Burlington Road\\
          Dublin 4, Ireland
        }
      \end{center}
      \begin{center}
        \texttt{Email: vbraun@stp.dias.ie}
      \end{center}
    \end{minipage}
  \end{center}
  \vspace*{\stretch1}
  \begin{abstract}
    Calabi-Yau threefolds with $h^{11}(X)=h^{21}(X)=1$ are constructed
    as free quotients of a hypersurface in the ambient toric variety
    defined by the 24-cell. Their fundamental groups are $SL(2,3)$,
    $\Z_3\rtimes \Z_8$, and $\Z_3\times Q_8$.
  \end{abstract}
  \vspace*{\stretch1}
\end{titlepage}
\tableofcontents
\listoffigures 	
\listoftables 	

\section{Introduction}
\label{sec:into}

\begin{figure}[bpt]
  \centering
  \includegraphics[width=13cm]{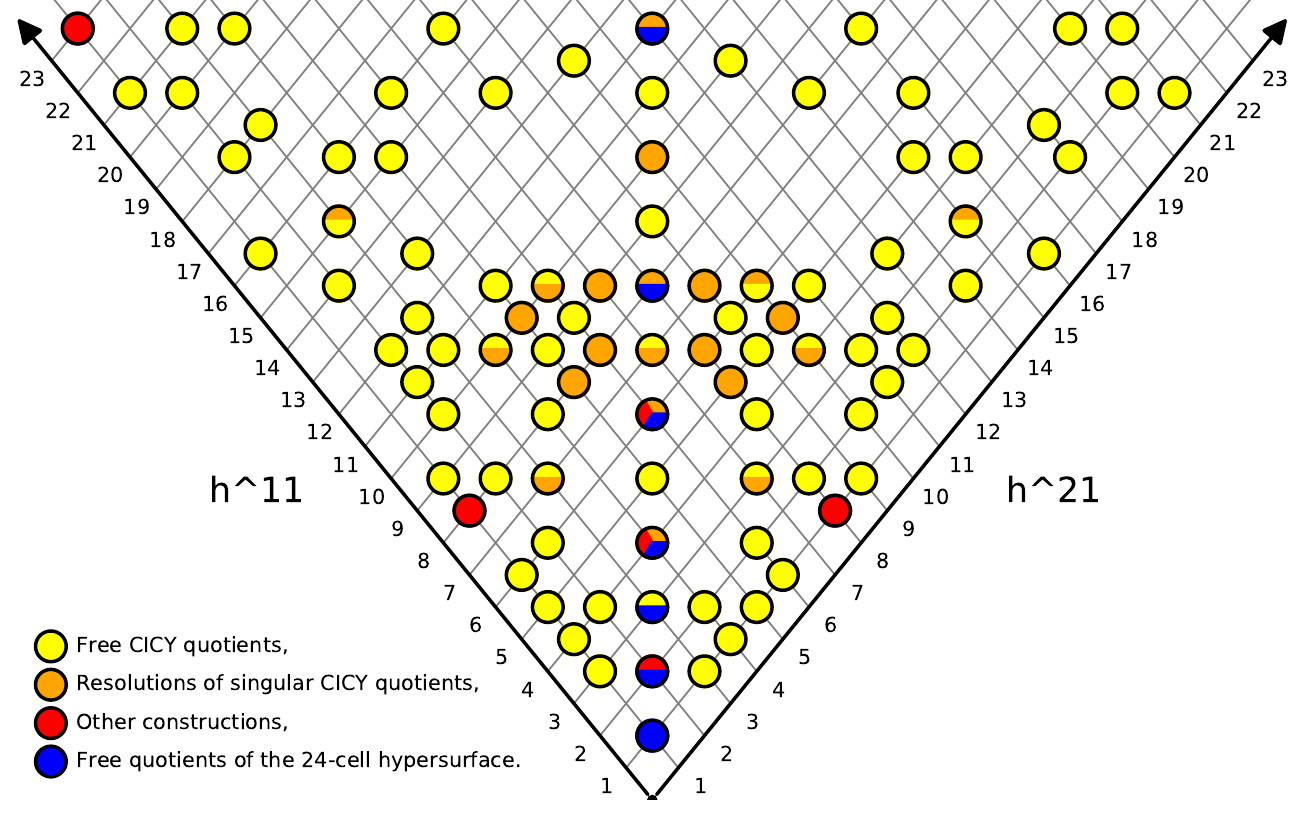}
  \caption{Plot of the Hodge numbers of known Calabi-Yau threefolds
    (and their mirrors) with $h^{11}+h^{21}<25$.}
  \label{fig:hodge25}
\end{figure}
The most basic observation about the Hodge numbers of Calabi-Yau
threefolds is that they apparently cannot take arbitrary values, even
though we do not have any good mathematical explanation. One empirical
constraint~\cite{Kreuzer:2000xy} is that the height is limited to
\begin{equation}
  h^{11} + h^{21} \leq 502
  ,
\end{equation}
leaving us only with a finite number of possible Hodge numbers. Not
having made any progress in the way of an upper bound for the height,
one might want to ask whether there is any lower
bound~\cite{Candelas:2007ac, Candelas:2008wb, Candelas:2010ve}. In
fact, just looking at the lists of complete intersections in
projective spaces (CICY~\cite{Candelas:1987kf, Candelas:1987du})
yields $h^{11}+h^{21}\geq 30$ and the hypersurfaces in toric varieties
satisfy $h^{11}+h^{21}\geq 29$. However, dividing out free group
actions almost always lowers the Hodge numbers (and never raises
them), so these naive lower bounds for the height can be easily
violated~\cite{Davies:2009ub, Davies:2011is, Filippini:2011rf}. From a
physics perspective, this serves both to reduce the number of
moduli~\cite{Braun:2004xv, Braun:2007tp, Braun:2007xh, Braun:2007vy}
and, via the Hosotani mechanism~\cite{Hosotani1, Hosotani2,
  WittenSBP}, to break the GUT gauge group. By systematically
constructing free quotients of CICY threefolds~\cite{gross-2005,
  Candelas:2008wb, GrossPopescu1089G, MR2373582, hua-2007,
  Braun:2010vc}, one can push down the lower boundary for the height
to $h^{11}+h^{21}=4$. In particular, a minimal three-generation
manifold~\cite{Braun:2009qy} with $(h^{11},h^{21})=(4,1)$ can thus be
realized.

However, one might wonder if even smaller Hodge numbers are
possible. In particular, the minimal value for a non-rigid Calabi-Yau
threefold would be $(h^{11},h^{21})=(1,1)$. The purpose of this paper
is to fill this gap, and construct a ``minimal'' Hodge number
example. The idea, in a nutshell, is to look for permutation actions
that act simply transitively~\footnote{That is, for any two vertices
  $v_1$, $v_2$ there exists a unique group element $g\in G$ with
  $g(v_1)=v_2$.} on the vertices of a lattice polytope, and use this
to define a group action on a anticanonical hypersurface in the
corresponding toric variety. Those familiar with such constructions
will immediately notice that this almost implies that there is a
single complex structure modulus. However, various technical details
need to be checked before one can conclude that this quotient is,
indeed, a smooth Calabi-Yau threefold.

In \autoref{sec:24}, I will start with some elementary properties of
the $24$-cell lattice polytope that I will use in the following. In
\autoref{sec:SL23}, I am going to define a group action and compute
the fixed point sets on the ambient toric variety defined by the
lattice polytope. Then, in~\autoref{sec:CY}, I will check that it
leads to a desired free action on a Calabi-Yau hypersurface, leading
to Hodge numbers $(1,1)$ on the smooth quotient threefold. Finally, in
\autoref{sec:general}, I will quickly go through two similar group
actions and all partial quotients. All toric geometry computations
used in this paper were done using~\cite{ToricVarieties, Sage, GPS05,
  GAP4}.

\section{The 24-Cell}
\label{sec:24}

There are six 4-dimensional regular polytopes, the 4-simplex, 4-cube,
16-cell, 24-cell, 120-cell and 600-cell. Apart from the 24-cell, these
are higher-dimensional analogues of the tetrahedron, cube, octahedron,
dodecahedron, and icosahedron. In particular, they transform in the
same way under duality.\footnote{The tetrahedron is self-dual, cube
  and octahedron are exchanged, and dodecahedron and icosahedron are
  exchanged.} The 24-cell is the regular 4-dimensional polytope that
does not have a 3-dimensional analog. For lack of anything else to
transform into, it is also self-dual. A curious fact, that has already
been remarked in~\cite{Kreuzer:2000xy}, is that the 24-cell appears as
one of the 473,800,776 reflexive 4-d lattice polytopes. In fact, the
24-cell can be constructed as the convex hull of the 24 roots of the
$D_4$ lattice. Amongst all 4-d reflexive lattice polytopes, it is the
one with the largest symmetry group~\cite{Kreuzer:2000xy}.
\begin{figure}
  \centering
  \includegraphics[width=15cm]{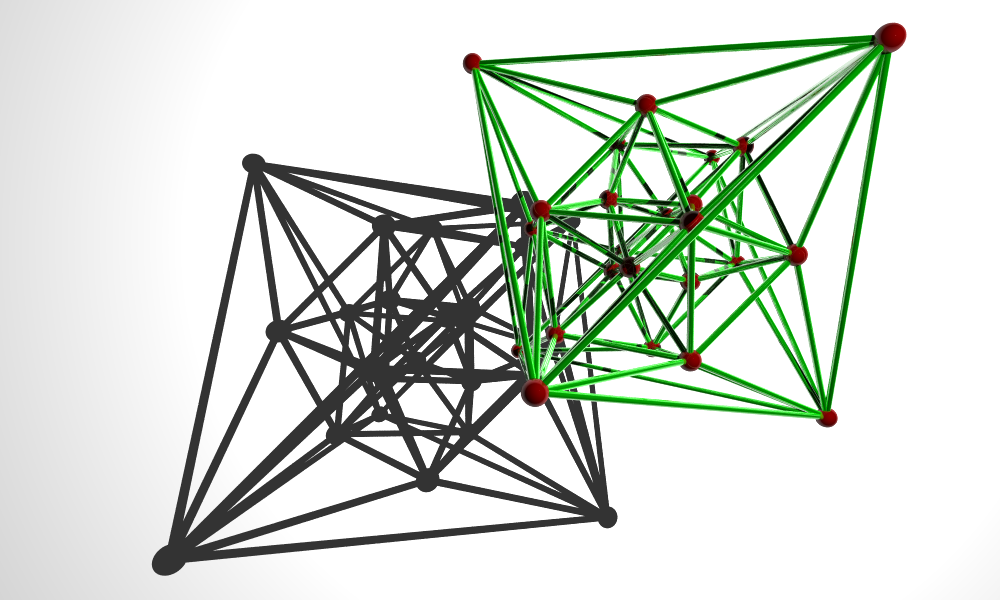}
  \caption{Steregraphic projection of the 24-cell into 3 dimensions.}
  \label{fig:24}
\end{figure}
The symmetry group obviously must contain the Weyl group of $D_4$, but
is actually larger. In fact, the full symmetry group of the 24-cell is
the Weyl group of $F_4$ and has $1152$ elements.

\section{A Toric Variety with SL(2,3) Action}
\label{sec:SL23}

\subsection{The Face Fan}
\label{sec:face}

Following the usual notation of toric geometry~\cite{Fulton,
  Batyrev:1994hm, CoxLittleSchenck}, we will identify the
4-dimensional root lattice of $D_4$ with $N\simeq \Z^4$. Doing so
breaks much of the symmetry and the coordinates of the vertices do not
manifest the 24-cell structure at all. However, picking a basis is
convenient for direct computation and we will the particular lattice
basis of Tables~\ref{tab:points} and~\ref{tab:facets} in the
following.
\begin{table}
  \centering
  \begin{sideways}
    \begin{tabular}{|c||r@{}r@{$,\,$}r@{$,\,$}r@{$,\,$}r@{}l|c|cc|c|}
\hline$n$ & & \multicolumn{4}{c}{$p_n$} & & $z_n$ & $g_3(n)$ & $g_4(n)$ & facets\\
\hline
  1 & $($ & 1 & 0 & 0 & 0 & $)$ & $z_{1}$ & 14 & 10 & STUVWX \\
  2 & $($ & 0 & 1 & 0 & 0 & $)$ & $z_{2}$ & 24 & 12 & GORSTW \\
  3 & $($ & 0 & 0 & 1 & 0 & $)$ & $z_{3}$ & 18 & 9 & EHNTUV \\
  4 & $($ & 0 & 0 & 0 & 1 & $)$ & $z_{4}$ & 10 & 19 & GHPSTU \\
  5 & $($ & 1 & -1 & -1 & 1 & $)$ & $z_{5}$ & 21 & 2 & DPQSUX \\
  6 & $($ & 0 & 0 & -1 & 1 & $)$ & $z_{6}$ & 12 & 20 & GIPQRS \\
  7 & $($ & 0 & -1 & 0 & 1 & $)$ & $z_{7}$ & 2 & 1 & DEHJPU \\
  8 & $($ & -1 & 0 & 0 & 1 & $)$ & $z_{8}$ & 19 & 3 & GHIJKP \\
  9 & $($ & 1 & 0 & 0 & -1 & $)$ & $z_{9}$ & 23 & 14 & ABCVWX \\
  10 & $($ & 0 & 1 & 0 & -1 & $)$ & $z_{10}$ & 15 & 16 & ABMORW \\
  11 & $($ & 0 & 0 & 1 & -1 & $)$ & $z_{11}$ & 5 & 13 & BCELNV \\
  12 & $($ & -1 & 1 & 1 & -1 & $)$ & $z_{12}$ & 17 & 15 & BKLMNO \\
  13 & $($ & 1 & -1 & -1 & 0 & $)$ & $z_{13}$ & 8 & 6 & ACDFQX \\
  14 & $($ & 0 & 0 & -1 & 0 & $)$ & $z_{14}$ & 22 & 8 & AFIMQR \\
  15 & $($ & 0 & -1 & 0 & 0 & $)$ & $z_{15}$ & 4 & 5 & CDEFJL \\
  16 & $($ & -1 & 0 & 0 & 0 & $)$ & $z_{16}$ & 3 & 7 & FIJKLM \\
  17 & $($ & 1 & -1 & 0 & 0 & $)$ & $z_{17}$ & 6 & 18 & CDEUVX \\
  18 & $($ & 1 & 0 & -1 & 0 & $)$ & $z_{18}$ & 16 & 21 & AQRSWX \\
  19 & $($ & 0 & 1 & 1 & -1 & $)$ & $z_{19}$ & 13 & 24 & BNOTVW \\
  20 & $($ & -1 & 1 & 1 & 0 & $)$ & $z_{20}$ & 9 & 11 & GHKNOT \\
  21 & $($ & -1 & 1 & 0 & 0 & $)$ & $z_{21}$ & 11 & 22 & GIKMOR \\
  22 & $($ & -1 & 0 & 1 & 0 & $)$ & $z_{22}$ & 1 & 17 & EHJKLN \\
  23 & $($ & 0 & -1 & -1 & 1 & $)$ & $z_{23}$ & 20 & 4 & DFIJPQ \\
  24 & $($ & 0 & 0 & 0 & -1 & $)$ & $z_{24}$ & 7 & 23 & ABCFLM \\
\hline
\end{tabular}

  \end{sideways}
  \caption[The vertices $p_n$ of the 24-cell lattice polytope.]{The
    vertices $p_n$ of the 24-cell lattice polytope. The $z_n$
    column is the corresponding homogeneous variable of the toric
    variety, the next two columns are the transformation under the
    group action in eq.~\eqref{eq:g3g4perm}, and the final column
    lists the incident facets labeled as in Table~\ref{tab:facets}.}
  \label{tab:points}
\end{table}
\begin{table}
  \centering
  \begin{sideways}
    \begin{tabular}{|c||r@{}r@{$,\,$}r@{$,\,$}r@{$,\,$}r@{}l@{$\cdot \vec{x} = $}r|}
\hline$N$ & & \multicolumn{5}{c}{equation} & \\
\hline
  A & $($ & 0 & 0 & -1 & -1 & $)$ & 1 \\
  B & $($ & 0 & 0 & 0 & -1 & $)$ & 1 \\
  C & $($ & 0 & -1 & 0 & -1 & $)$ & 1 \\
  D & $($ & 0 & -1 & 0 & 0 & $)$ & 1 \\
  E & $($ & 0 & -1 & 1 & 0 & $)$ & 1 \\
  F & $($ & -1 & -1 & -1 & -1 & $)$ & 1 \\
  G & $($ & 0 & 1 & 0 & 1 & $)$ & 1 \\
  H & $($ & 0 & 0 & 1 & 1 & $)$ & 1 \\
  I & $($ & -1 & 0 & -1 & 0 & $)$ & 1 \\
  J & $($ & -1 & -1 & 0 & 0 & $)$ & 1 \\
  K & $($ & -1 & 0 & 0 & 0 & $)$ & 1 \\
  L & $($ & -1 & -1 & 0 & -1 & $)$ & 1 \\
  M & $($ & -1 & 0 & -1 & -1 & $)$ & 1 \\
  N & $($ & 0 & 0 & 1 & 0 & $)$ & 1 \\
  O & $($ & 0 & 1 & 0 & 0 & $)$ & 1 \\
  P & $($ & 0 & 0 & 0 & 1 & $)$ & 1 \\
  Q & $($ & 0 & 0 & -1 & 0 & $)$ & 1 \\
  R & $($ & 0 & 1 & -1 & 0 & $)$ & 1 \\
  S & $($ & 1 & 1 & 0 & 1 & $)$ & 1 \\
  T & $($ & 1 & 1 & 1 & 1 & $)$ & 1 \\
  U & $($ & 1 & 0 & 1 & 1 & $)$ & 1 \\
  V & $($ & 1 & 0 & 1 & 0 & $)$ & 1 \\
  W & $($ & 1 & 1 & 0 & 0 & $)$ & 1 \\
  X & $($ & 1 & 0 & 0 & 0 & $)$ & 1 \\
\hline
\end{tabular}

  \end{sideways}
  \caption{The facets of the 24-cell lattice polytope.}
  \label{tab:facets}
\end{table}
Given these $24$ points, we define the polytope
\begin{equation}
  \nabla = \conv\big\{ p_1,\, p_2,\, \dots,\, p_{24} \big\}
  .
\end{equation}
Each of the $24$ facets of $\lambda$ is an octahedron, and spans one
generating cone in the face fan 
\begin{multline}
  \fan = 
  \scriptstyle
  \big\{
  \langle p_{2},\,p_{6},\,p_{10},\,p_{14},\,p_{18},\,p_{21}\rangle,~
  \langle p_{10},\,p_{12},\,p_{14},\,p_{16},\,p_{21},\,p_{24}\rangle,~
  \langle p_{2},\,p_{10},\,p_{12},\,p_{19},\,p_{20},\,p_{21}\rangle,
  \\
  \scriptstyle
  \langle p_{6},\,p_{8},\,p_{14},\,p_{16},\,p_{21},\,p_{23}\rangle,~
  \langle p_{2},\,p_{4},\,p_{6},\,p_{8},\,p_{20},\,p_{21}\rangle,~
  \langle p_{8},\,p_{12},\,p_{16},\,p_{20},\,p_{21},\,p_{22}\rangle,
  \langle p_{1},\,p_{5},\,p_{9},\,p_{13},\,p_{17},\,p_{18}\rangle,~
  \\
  \scriptstyle
  \langle p_{9},\,p_{11},\,p_{13},\,p_{15},\,p_{17},\,p_{24}\rangle,~
  \langle p_{1},\,p_{3},\,p_{9},\,p_{11},\,p_{17},\,p_{19}\rangle,
  \langle p_{5},\,p_{7},\,p_{13},\,p_{15},\,p_{17},\,p_{23}\rangle,~
  \langle p_{1},\,p_{3},\,p_{4},\,p_{5},\,p_{7},\,p_{17}\rangle,~
  \\
  \scriptstyle
  \langle p_{3},\,p_{7},\,p_{11},\,p_{15},\,p_{17},\,p_{22}\rangle,
  \langle p_{9},\,p_{10},\,p_{13},\,p_{14},\,p_{18},\,p_{24}\rangle,~
  \langle p_{1},\,p_{2},\,p_{9},\,p_{10},\,p_{18},\,p_{19}\rangle,~
  \langle p_{9},\,p_{10},\,p_{11},\,p_{12},\,p_{19},\,p_{24}\rangle,
  \\
  \scriptstyle
  \langle p_{5},\,p_{6},\,p_{13},\,p_{14},\,p_{18},\,p_{23}\rangle,~
  \langle p_{1},\,p_{2},\,p_{4},\,p_{5},\,p_{6},\,p_{18}\rangle,~
  \langle p_{4},\,p_{5},\,p_{6},\,p_{7},\,p_{8},\,p_{23}\rangle,
  \langle p_{13},\,p_{14},\,p_{15},\,p_{16},\,p_{23},\,p_{24}\rangle,~
  \\
  \scriptstyle
  \langle p_{1},\,p_{2},\,p_{3},\,p_{4},\,p_{19},\,p_{20}\rangle,~
  \langle p_{11},\,p_{12},\,p_{15},\,p_{16},\,p_{22},\,p_{24}\rangle,
  \langle p_{3},\,p_{11},\,p_{12},\,p_{19},\,p_{20},\,p_{22}\rangle,~
  \langle p_{7},\,p_{8},\,p_{15},\,p_{16},\,p_{22},\,p_{23}\rangle,~
  \\
  \scriptstyle
  \langle p_{3},\,p_{4},\,p_{7},\,p_{8},\,p_{20},\,p_{22}\rangle
  \displaystyle
  \big\}.
\end{multline}

\begin{figure}
  \centering
  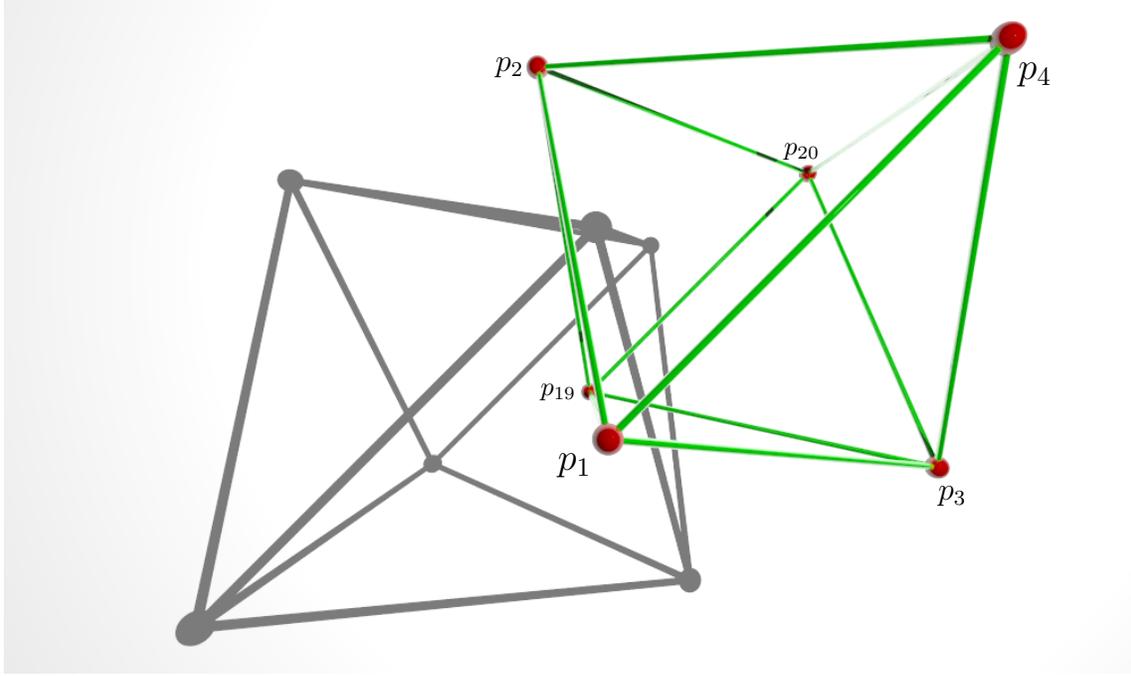
  \caption{One facet of the 24-cell with its spanning vertices.}
  \label{fig:octahedron}
\end{figure}

We will now pick a particular $24$-element subgroup of the
automorphism group $\mathop{\mathrm{Weyl}}(F_4)$ of the
24-cell. Acting on from the left on the vertices $p_i \in N$ of the
polytope $\nabla$, it is generated by the two matrices
\begin{equation}
    g_3 =
    \begin{pmatrix}
      0 & 0 & 1 & 0 \\
      0 & 0 & 0 & 1 \\
      -1 & 0 & -1 & 0 \\
      0 & -1 & 0 & -1
    \end{pmatrix}
    ,\quad
    \quad
    g_4 =
    \begin{pmatrix}
      0 & -1 & 1 & 0 \\
      1 & 1 & 0 & 1 \\
      0 & 1 & 0 & 1 \\ 
      -1 & -1 & -1 & -1
    \end{pmatrix}
    .
    \label{eq:g3g4def}
\end{equation}
Alternatively, the two generators can be written as the two
permutations
\begin{equation}
  \label{eq:g3g4perm}
  \begin{split}
    g_3 =&\;
    \text{
      \small
      $(1,14,22)(2,24,7)(3,18,16)(4,10,15)(5,21,11)(6,12,17)(8,19,13)(9,23,20)$
    }
    \\
    g_4 =&\;
    \text{
      \small
      $(1,10,16,7)(2,12,15,5)(3,9,14,8)(4,19,24,23)(6,20,11,13)(17,18,21,22)$
    }
  \end{split}
\end{equation}
permuting the $24$ vertices\footnote{By abuse of notation, we will not
  distinguish the group from its matrix and permutation representation
  in the following.}. Together, $g_3$ and $g_4$ generate a
representation of the group
\begin{equation}
  G 
  \eqdef
  \langle g_3,\, g_4 \rangle 
  \simeq
  SL(2,3),
\end{equation}
of $2\times 2$-matrices with entries in the finite field with $3$
elements. Using the permutation group action, we can write the
complete fan (including all lower-dimensional faces) succinctly as
\begin{equation}
  \begin{split}
    \fan =&\;
    G \cdot \big\{ 
    \langle p_{1},\,p_{2},\,p_{3},\,p_{4},\,p_{19},\,p_{20}\rangle \big\} 
    ~ \cup
    \\ &\;
    G \cdot \big\{ 
    \langle p_{1},\,p_{2},\,p_{4} \rangle,~
    \langle p_{1},\,p_{3},\,p_{4} \rangle,~ 
    \langle p_{1},\,p_{2},\,p_{19} \rangle,~
    \langle p_{1},\,p_{3},\,p_{19} \rangle 
    \big\} ~ \cup
    \\ &\;
    G \cdot \big\{ 
    \langle p_{1},\,p_{2} \rangle,~
    \langle p_{1},\,p_{3} \rangle,~ 
    \langle p_{1},\,p_{4} \rangle,~
    \langle p_{1},\,p_{19} \rangle 
    \big\} ~ \cup
    \\ &\;
    G \cdot \big\{ \langle p_1 \rangle \big\} ~\cup~
    \big\{ \langle\rangle \big\}
  \end{split}
\end{equation}
See also \autoref{fig:octahedron} for a pictorial representation of
the relative position of the rays of the cones.

\subsection{Homogeneous Coordinates and the Maximal Torus}
\label{sec:torus}

In the following, I will be using the Cox homogeneous coordinate
description of the toric variety~\cite{MR1299003}. There are $24$
homogeneous coordinates $z_1$, $\dots$, $z_{24}$ modulo
\begin{equation}
  \Hom\big( A_3(\CP_\nabla), \C^\times \big) = \big(\C^\times)^{20}
\end{equation}
rescalings. By definition, the 4-dimensional toric variety
$\CP_\nabla$ comes with an action of $(\C^\times)^4$ such that there
is a single maximal-dimensional orbit, which I will denote as
$\CP_\emptycone$ in the following. In the usual correspondence between
torus orbits and cones of the fan, this is the orbit associated to the
trivial cone $\langle\rangle$. In terms of homogeneous coordinates, it
is the locus where no homogeneous coordinate vanishes. The maximal
orbit itself is always smooth as singular points must fill out whole
torus orbits.

In the remainder of this section, we will be investigating the fixed
points of $G\simeq SL(2,3)$ on the maximal torus. Since all
homogeneous coordinates are invertible there, we can freely use the
homogeneous rescalings to set some homogeneous coordinates to
unity. In fact, we can pick unique representatives
\begin{equation}
  \begin{split}
    \CP_\emptycone \;&=
    \Big\{
    [\hat z_1:\cdots:\hat z_{24}] 
    ~\Big|~
    \hat z_i\not=0
    \Big\}
    = 
    \\&=
    \Big\{
    [z_1:z_2:z_3:z_4:1:\cdots:1]
    ~\Big|~
    z_i\not=0
    \Big\}
    \simeq \big(\C^\times\big)^4
    \subset P_\nabla
  \end{split}
\end{equation}
for each point where only the first four homogeneous coordinates are
non-zero, and such that there are no remaining identifications. This
is so because there is the following \emph{integral basis} for the
linear relations amongst the vertices of $\nabla$:
\begin{equation}
  \begin{array}{c@{~=~}r@{}r@{}r@{}r}
   p_{ 5} &  p_1 & -p_2 & -p_3 & +p_4 \\
   p_{ 6} &      &      & -p_3 & +p_4 \\
   p_{ 7} &      & -p_2 &      & +p_4 \\
   p_{ 8} & -p_1 &      &      & +p_4 \\
   p_{ 9} &  p_1 &      &      & -p_4 \\
   p_{10} &      &  p_2 &      & -p_4 \\
   p_{11} &      &      &  p_3 & -p_4 \\
   p_{12} & -p_1 & +p_2 & +p_3 & -p_4 \\
   p_{13} &  p_1 & -p_2 & -p_3 &   \\
   p_{14} &      &      & -p_3 &   \\
  \end{array}
  \hspace{2cm}
  \begin{array}{c@{~=~}r@{}r@{}r@{}r}
   p_{15} &      & -p_2 &      &   \\
   p_{16} & -p_1 &      &      &   \\
   p_{17} &  p_1 & -p_2 &      &   \\
   p_{18} &  p_1 &      & -p_3 &   \\
   p_{19} &      &  p_2 & +p_3 & -p_4 \\
   p_{20} & -p_1 & +p_2 & +p_3 &   \\
   p_{21} & -p_1 & +p_2 &      &   \\
   p_{22} & -p_1 &      & +p_3 &   \\
   p_{23} &      & -p_2 & -p_3 & +p_4 \\
   p_{24} &      &      &      & -p_4
  \end{array}
  \label{eq:relsbasis}
\end{equation}
The first equation (for $p_5$) then translates into the homogeneous
rescaling
\begin{equation}
  \big[ z_1:\cdots:z_{24} \big] = 
  \big[
  \lambda z_1 :
  \lambda^{-1} z_2 :
  \lambda^{-1} z_3 :
  \lambda z_4 :
  \lambda^{-1} z_5 :
  z_6:z_7:\cdots:z_{24} 
  \big]
\end{equation}
and so on. Clearly, the basis for the relations
eq.~\eqref{eq:relsbasis} allows us to unambiguously scale $z_5$,
$\dots$, $z_{24}$ to unity on the maximal torus.

\subsection{Fixed Points on the Maximal Torus}
\label{sec:torusfp}

To find the fixed points of the $G$-action on the toric variety
$\CP_\nabla$, we need to look at each conjugacy class of
$G$. Excluding the identity of $G$, there are $6$ non-trivial
conjugacy classes. For the remainder of this section, I will discuss
each in turn.

\subsubsection*{The conjugacy class of \boldmath$g_4^2$}

The first conjugacy class is of order $2$ and contains the single
central group element
\begin{equation}
  g_4^2 = \diag(-1,-1,-1,-1).
\end{equation}
Its action on the homogeneous coordinates on the maximal torus
$\CP_\emptycone$ is
\begin{multline}
  g_4^2
  \Big(\big[ z_1:z_2:z_3:z_4:1:\cdots:1 \big]\Big)
  =
  \\
  =
  \big[ 
  1\!:\!\cdots\!:\!1:
  \underbrace{z_3}_{\text{Position}~14}:
  \underbrace{z_2}_{15}:
  \underbrace{z_1}_{16}:
  1\!:\!\cdots\!:\!1:
  \underbrace{z_4}_{\text{Position}~24}
  \big]
  =
  \\
  =
  \left[
    \frac{1}{z_1}:\frac{1}{z_2}:\frac{1}{z_3}:\frac{1}{z_4}:1:\cdots:1
  \right]
  .
\end{multline}
Hence, there are $2^4$ fixed points 
\begin{equation}
  \CP_\emptycone^{g_4^2}
  =
  \Big\{
  \big[ z_1:z_2:z_3:z_4:1:\cdots:1 \big]
  ~\Big|~
  z_0,z_2,z_2,z_3 \in \{+1, -1\}
  \Big\}
\end{equation}

\subsubsection*{The conjugacy class of \boldmath$g_4$}

The second conjugacy class is of order $4$ and contains $6$ group
elements. Since a $g_4$-fixed point is also a $g_4^2$-fixed point, we
immediately note that $\CP_\emptycone^{g_4} \subset
\CP_\emptycone^{g_4^2}$ is again a discrete set of points. Explicitly,
the $g_4$-fixed point set turns out to be $4$ out of the $12$ fixed
points of $g_4^2$, namely
\begin{equation}
  \begin{split}
    \CP_\emptycone^{g_4} 
    = 
    \smash{\Big\{}~
    &
    [ +1:+1:+1:+1: 1:\cdots:1 ]   ,\\&
    [ +1:-1:-1:+1: 1:\cdots:1 ]   ,\\&
    [ -1:+1:-1:-1: 1:\cdots:1 ]   ,\\&
    [ -1:-1:+1:-1: 1:\cdots:1 ]
    ~\smash{\Big\}}
    .
  \end{split}
\end{equation}

\subsubsection*{The conjugacy classes of \boldmath$g_3$ and $g_3^2$}

The third and fourth conjugacy class can be represented by $g_3$ and
$g_3^2$, respectively. They are both of order $3$ and contain $4$
representatives. In fact, they are related by an automorphism of $G$
and, therefore, one only needs to discuss $g_3$, say. On the maximal
torus $\CP_\emptycone$, its action is
\begin{equation}
  g_3
  \Big(\big[ z_1:z_2:z_3:z_4:1:\cdots:1 \big]\Big)
  = 
  \left[ 
    z_3:
    z_4:
    \frac{1}{z_1 z_3}:
    \frac{1}{z_2 z_4}:
    1:\cdots:1 
  \right]
\end{equation}
and, therefore, the fixed points are given by the ideal\footnote{I
  will always think of the equations of the fixed points on the
  maximal torus as an ideal in the polynomial ring in the formal
  commutative variables $z_1$, $z_1^{-1}$, $\dots$, $z_4$, $z_4^{-1}$
  subject to the relations $z_1 z_1^{-1}=1$, $\dots$, $z_4
  z_4^{-1}=1$.}
\begin{equation}
  \begin{split}
    I^{g_3} = 
    \left<
      z_1 = z_3,~
      z_2 = z_4,~
      z_3 = \frac{1}{z_1 z_3},~
      z_4 = \frac{1}{z_2 z_4}
    \right>
  \end{split}
  .
\end{equation}
The associated variety $V(I)$ is the fixed point set, and an
elementary computation reveals that it consists of the $9$ points
\begin{equation}
  \CP_\emptycone^{g_3} 
  =
  V\big( I^{g_3} \big)
  =
  \Big\{
  \big[ \mu:\nu:\mu:\nu:1:\cdots:1 \big]
  ~\Big|~
  \mu, \nu \in \{ 1, e^{2\pi i/3}, e^{4\pi i/3}\}
  \Big\}
  .
\end{equation}

\subsubsection*{The conjugacy classes of \boldmath$g_3 g_4^2$ and $g_3^2 g_4^2$}

The fifth and sixth conjugacy class can be represented by $g_3g_4^2$
and $g_3^2g_4^2$, respectively. They are both of order $6$ and contain
$4$ representatives. These are again related by an automorphism of $G$
and, therefore, one only needs to discuss
\begin{equation}
  g_3 g_4^2 
  =
  \begin{pmatrix}
    0 & 0 & -1 & 0 \\
    0 & 0 & 0 & -1 \\
    1 & 0 & 1 & 0 \\
    0 & 1 & 0 & 1
  \end{pmatrix}
  .
\end{equation}
Moreover, $(g_3g_4^2)^2=g_3^2$, so the fixed point set of $g_3g_4^2$
is contained in the $9$ fixed points of $g_3^2$. An explicit
computation shows that $g_3g_4^2$ fixes the single point
\begin{equation}
  \CP_\emptycone^{g_3 g_4^2} 
  = 
  \Big\{
  \big[ 1:1:1:1:1:\cdots:1 \big]
  \Big\}
\end{equation}
on the maximal torus.

\subsection{Stanley-Reisner Ideal and Other Fixed Points}
\label{sec:otherfp}

Thus far, I have shown that any element of $G\simeq SL(2,3)$ fixes a
discrete set of points on the maximal torus $\CP_\emptycone\simeq
(\C^\times)^4$, that is, on the locus where all homogeneous variables
are non-zero. The $4$-dimensional toric variety $\CP_\nabla$ is a
compactification of $\CP_\emptycone$ by gluing in lower-dimensional
toric varieties associated to the non-empty cones of the fan
$\fan$. One needs to discuss fixed points on these lower-dimensional
strata as well. The main observation is that, given any $g\in G$, if
the $i$-th homogeneous variable $z_i$ vanishes, the permuted
homogeneous variable $z_{g(i)}$ has to vanish as well on the $g$-fixed
point set. In other words, the $g$-fixed point sets on the
lower-dimensional strata are contained in the subvarieties where whole
$g$-permutation orbits of homogeneous variables vanish. As in the
previous section, I will discuss each conjugacy class separately. Note
that because $G$ acts regularly (simply transitively) on the $24$
variables, all $g\in G$-orbits are of the same size.

\subsubsection*{The conjugacy class of \boldmath$g_4^2$}

The $24$ homogeneous variables $z_1$, $\dots$, $z_{24}$ form $12$
orbits of length $2$ under the permutation $g_4^2$. For example, the
orbit containing the first variable is $\{z_1, z_{g_4^2(i)}\} =
\{z_1,z_{16}\}$. Therefore, the fixed point set away from the maximal
torus is contained in the union of the $12$ subvarieties
\begin{equation}
  \label{eq:FpOutg42}
  \Big( \CP_\nabla - \CP_\emptycone \Big)^{g_4^2}
  ~\subset~
  \bigcup_{g\in G}
  V\big(\langle z_{g(1)}=0,~ z_{g(16)}=0 \rangle\big)
  .
\end{equation}

\subsubsection*{The conjugacy class of \boldmath$g_4$}

The homogeneous variables form $6$ orbits of length $4$. The fixed
point set away from the maximal torus is 
\begin{equation}
  \label{eq:FpOutg4}
  \Big( \CP_\nabla - \CP_\emptycone \Big)^{g_4}
  ~\subset~
  \bigcup_{g\in G}
  V\big(\langle 
  z_{g(1)}=0,~ 
  z_{g(7)}=0,~ 
  z_{g(10)}=0,~ 
  z_{g(16)}=0 
  \rangle\big)
  .
\end{equation}

\subsubsection*{The conjugacy classes of \boldmath$g_3$ and $g_3^2$}

The homogeneous variables form $8$ orbits of length $3$. Again, it
suffices to consider $g_3$-fixed points. Away from the maximal torus,
they are
\begin{equation}
  \label{eq:FpOutg3}
  \Big( \CP_\nabla - \CP_\emptycone \Big)^{g_3}
  ~\subset~
  \bigcup_{g\in G}
  V\big(\langle 
  z_{g(1)}=0,~ 
  z_{g(14)}=0,~ 
  z_{g(22)}=0 
  \rangle\big)
  .
\end{equation}

\subsubsection*{The conjugacy classes of \boldmath$g_3 g_4^2$ and
  $g_3^2 g_4^2$}

The homogeneous variables form $4$ orbits of length $6$. Again, it
suffices to consider $g_3 g_4^2$-fixed points. Away from the maximal
torus, they are
\begin{multline}
  \label{eq:FpOutg3g42}
  \Big( \CP_\nabla - \CP_\emptycone \Big)^{g_3 g_4^2}
  ~\subset~
  \bigcup_{g\in G}
  V\big(\langle 
  z_{g(1)}=0,~ 
  z_{g(3)}=0,~ 
  z_{g(14)}=0,~ 
  \\
  z_{g(16)}=0,~ 
  z_{g(18)}=0,~ 
  z_{g(22)}=0 
  \rangle\big)
  .
\end{multline}

\subsubsection*{The Stanley-Reisner Ideal}

Not all homogeneous variables are allowed to vanish simultaneously as
one can see from the homogeneous coordinate description
\begin{equation}
  \CP_\nabla = 
  \frac{
    \C^{\fan(1)} - Z(\fan)
  }{
    (\C^\times)^{20}
  }
\end{equation}
of the toric variety. Here, $\fan(1)$ are the $24$ rays of the fan,
$\C^{\fan(1)}\simeq \C^{24}$ is the affine space parametrized by the
corresponding homogeneous variables, and $Z(\fan)$ is the exceptional
set
\begin{equation}
  Z(\fan)
  = 
  \bigcup_{z_{i_1} \cdots z_{i_k} \in SR(\fan)}
  V\big(
  z_{i_1}=0,~
  \dots,~
  z_{i_k}=0
  \big)
  .
\end{equation}
Taking the union over all monomials in the
Stanley-Reisner ideal $SR(\fan)$ is equivalent to only taking the
union over the finitely many minimal monomials\footnote{The
  ``minimal'' monomials are $z_{i_1}\cdots z_{i_k}$ such that
  $\{i_1,\;\dots,\;i_k\}$ is a primitive collection.}

The Stanley-Reisner ideal of the variety $\CP_\nabla$ is generated by
$204$ monomials. Using the permutation group action, one can write it
as
\begin{multline}
  SR(\fan) = 
  \Big< \Big\{
  z_{g(1)} z_{g(16)},~
  z_{g(1)} z_{g(15)},~
  z_{g(1)} z_{g(14)},~
  z_{g(1)} z_{g(12)},~
  z_{g(1)} z_{g(24)},~
  \\
  z_{g(1)} z_{g(6)} z_{g(20)},~
  z_{g(1)} z_{g(6)} z_{g(10)},~
  z_{g(1)} z_{g(6)} z_{g(7)},~
  z_{g(1)} z_{g(6)} z_{g(13)}
  ~\Big|~
  g\in G
  \Big\}\Big>
  .
\end{multline}
We notice that any potential fixed point in
$\CP_\nabla-\CP_\emptycone$, that is, outside of the maximal torus, is
contained in the exceptional set, see eqns.~\eqref{eq:FpOutg42},
\eqref{eq:FpOutg4}, \eqref{eq:FpOutg3}, and~\eqref{eq:FpOutg3g42}. To
summarize, all fixed point sets of all non-trivial group elements
$g\in G$ are finite\footnote{Note that the fixed point set is
  automatically finite if it is contained in the maximal torus.} and
contained in the maximal torus $\CP_\emptycone$,
\begin{equation}
  \bigcup_{g\in G-\{1\}}
  \CP_\nabla^g
  ~\subset~
  \CP_\emptycone
  \simeq 
  (\C^\times)^4
  ~\subsetneq~
  \CP_\nabla
\end{equation}

\subsection{Singularities}
\label{sec:sing}

Each of the $24$ generating cones of the fan, for example $\langle
p_{1},\,p_{2},\,p_{3},\,p_{4},\,p_{19},\,p_{20}\rangle$, is not
smooth.\footnote{A cone $\sigma$ is smooth if the rays of the cone are
  a lattice basis for $N\cap \Span_\Q(\sigma)$. In other words, the
  associated open torus orbit is a smooth subset of the toric
  variety.} This gives rise to $24$ singular points of the toric
variety $\CP_\nabla$. All other cones of the fan (of dimension $\leq
3$) are smooth, and, therefore, $\CP_\nabla$ is smooth outside of the
$24$ singular points. 

In terms of homogeneous coordinates, these singular points are
\begin{equation}
  \label{eq:SingP}
  \Sing(\CP_\nabla) 
  =
  \Big\{
  V\big(z_{g(1)},~z_{g(2)},~z_{g(3)},~z_{g(4)},~z_{g(19)},~z_{g(20)}\big)
  ~\Big|~
  g \in G
  \Big\}
  .
\end{equation}
Since the maximal cones are not simplicial, the singular points are
worse than orbifold singularities by a finite group.

In the remainder of this section I will investigate the singularities
further. However, the details will not be important for the following.
Now, since each singularity is purely local data, it is most
convenient to use the description of the toric variety as patched
local affine schemes $\Spec\big( \C[\sigma^\dual \cap M]\big)$ instead
of the global description via homogeneous coordinates. By symmetry, I
just have to consider one of the $24$ affine patches, and will pick
\begin{equation}
  \begin{split}
    \sigma
    =
    \langle &p_{1},\,p_{2},\,p_{3},\,p_{4},\,p_{19},\,p_{20}\rangle
    \\
    \Leftrightarrow\quad
    \sigma^\dual 
    =
    \big<&
    (1,1,0,1),~
    (1,1,0,0),~
    (1,0,1,0),~
    (1,0,1,1),~
    \\&
    (0,0,1,1),~
    (0,1,0,1),~
    (0,1,0,0),~
    (0,0,1,0)
    \big>
  \end{split}
\end{equation}
Since the dual cone $\sigma^\dual$ is spanned by $8$ rays, we need an
$8$-dimensional ambient affine space to embed the patch $\Spec
\big(\C[\sigma^\dual \cap M]\big)$ of $\CP_\nabla$. A standard
computation~\cite{Hosten95grin:an, ToricVarieties} yields the toric
ideal
\begin{multline}
  \label{eq:Spec}
  \C\big[\sigma^\dual \cap M\big]
  = 
  \C\big[x_1,x_2,x_3,x_4,x_5,x_6,x_7,x_8\big]
  \Big/
  \\
  \big<
  x_1 x_3 - x_2 x_4,~
  x_1 x_5 - x_4 x_6,~
  x_1 x_7 - x_2 x_6,~
  x_1 x_8 - x_2 x_5,~
  x_2 x_5 - x_3 x_6,~
  \\
  x_2 x_5 - x_4 x_7,~
  x_2 x_8 - x_3 x_7,~
  x_3 x_5 - x_4 x_8,~
  x_5 x_7 - x_6 x_8
  \big>
\end{multline}
Note that the $9$ defining equations are far from transverse and, in
fact, cut out a $4$-dimensional affine algebraic variety in
$\C^8$. The singularity is at the origin $x_1=\cdots=x_8=0$.

\section{The Calabi-Yau Threefold}
\label{sec:CY}

\subsection{Construction and Smoothness}
\label{sec:Smooth}

We now pick a $G\simeq SL(2,3)$-invariant section of the
anti-canonical bundle on toric variety $\CP_\nabla$, yielding a
$3$-dimensional variety $\Xt$ with vanishing first Chern class
$c_1(\Xt)=0$. The sections of the anti-canonical bundle $-K_\nabla$ of
$\CP_\nabla$ are generated by the homogeneous monomials that
correspond to the integral points of the dual polytope $\Delta =
\nabla^\dual$, which are the origin and the $24$ vertices of $\Delta$
corresponding to the $24$ facets of $\nabla$. That
is~\cite{ToricVarieties},
\begin{multline}
  H^0\big(\CP_\nabla, -K_\nabla\big)
  =
  \Big<
  \Big\{ \prod_{i=1}^{24} z_i \Big\}
  \cup
  \Big\{
  z_{g(1)}^2 z_{g(2)}^2 z_{g(3)}^2 z_{g(4)}^2 
  z_{g(5)} z_{g(6)} z_{g(7)} z_{g(8)} z_{g(9)} 
  \times \\
  z_{g(10)} z_{g(11)} z_{g(12)} z_{g(17)} z_{g(18)} z_{g(19)}^2 
  z_{g(20)}^2 z_{g(21)} z_{g(22)}
  ~\Big|~
  g \in G
  \Big\}\Big>
\end{multline}
The permutation action of $G$ is facet-transitive on $\nabla$ and,
therefore, vertex-transitive on $\Delta$. Hence, the two invariant
polynomials are the monomial corresponding to the origin,
\begin{equation}
  \label{eq:DefPinfty}
  P_\infty 
  \eqdef
  \prod_{i=1}^{24} z_i 
  ,
\end{equation}
and the sum over the $24$ vertices of $\Delta$,
\begin{multline}
  \label{eq:DefP0}
  P_0 
  \eqdef
  \smash{
    \sum_{g\in G}
    \big(
  }
  z_{g(1)}^2 z_{g(2)}^2 z_{g(3)}^2 z_{g(4)}^2 
  z_{g(5)} z_{g(6)} z_{g(7)} z_{g(8)} z_{g(9)} 
  \times \\
  z_{g(10)} z_{g(11)} z_{g(12)} z_{g(17)} z_{g(18)} z_{g(19)}^2 
  z_{g(20)}^2 z_{g(21)} z_{g(22)}
  \smash{\big)}
\end{multline}
Together, they span the invariant sections
\begin{equation}
  \label{eq:aKinv}
  H^0\big(\CP_\nabla, -K_\nabla\big)^G
  =
  \Span_\C(P_0,~ P_\infty)
  .
\end{equation}
Hence, there is a one-parameter family
\begin{equation}
  \label{eq:Ptdef}
  P_\varphi \eqdef P_0 + \varphi P_\infty
  ,\quad
  \varphi \in \C\cup\{\infty\}
\end{equation}
of invariant polynomials, giving rise to a family
\begin{equation}
  \Xt_\varphi
  \eqdef
  \Big\{ P_\varphi = 0 \Big\} \subset 
  \CP_\nabla
\end{equation}
of $G\simeq SL(2,3)$-symmetric varieties. The quotient $X_\varphi=\Xt_\varphi/G$
is then again a $3$-dimensional variety with vanishing first Chern
class $c_1(X)=0$. There are $3$ potential sources for singularities on
the quotient, namely
\begin{enumerate}
\item singularities of the ambient toric variety $\CP_\nabla$
  containing $\Xt$,
\item fixed points of the $G$-action on $\Xt$, and
\item loci where the hypersurface equation $P_\varphi$ fails to be
  transverse.
\end{enumerate}
I will now discuss each case in turn.
\begin{enumerate}
\item Since $G$ permutes the $24$ singularities of $\CP_\nabla$, one
  only has to consider one of them. For example, the fixed point
  corresponding to the cone $\langle
  p_{1},\,p_{2},\,p_{3},\,p_{4},\,p_{19},\,p_{20}\rangle \in \fan$ is
  \begin{equation}
    s = 
    \big[0:0:0:0:1:\cdots:1:
    \underbrace{0:0}_{\mathclap{\text{Positions 19 and 20}}}
    :1:\cdots:1]
    \in \Sing\big(\CP_\nabla\big)
    ,
  \end{equation}
  see eq.~\eqref{eq:SingP}. At the singular point,
  \begin{equation}
    P_0(s) = 1
    ,\quad
    P_\infty(s) = 0
    .
  \end{equation}
  Therefore, a sufficiently generic hypersurface $\Xt_\varphi$ misses the
  singular point $s$.
\item On the maximal torus $\CP_\emptycone$ none of the homogeneous
  coordinates is allowed to vanish. In particular,
  \begin{equation}
    P_\infty(z) \not= 0 
    \quad \forall z\in \CP_\emptycone
  \end{equation}
  by eq.~\eqref{eq:DefPinfty}. Since all fixed points are contained in
  $\CP_\emptycone$, a sufficiently generic hypersurface $\Xt_\varphi$ misses
  the fixed points. For example, one can check that $\Xt_1$ misses all
  fixed points.
\item Finally, one has to check that the hypersurface equation is
  transverse. As in the end of \autoref{sec:sing}, I will make use of
  the covering of the toric variety $\CP_\nabla$ by $24$ affine
  patches $\Spec\big(\C[\sigma^\dual \cap M] \big)$ corresponding to
  the generating cones. By symmetry, one only has to check
  transversality in one of the $24$ affine patches. In the patch used
  for eq.~\eqref{eq:Spec}, the two invariant polynomials read
  \begin{equation}
    \begin{split}
      P_0\big|_{\Spec \C[\sigma^\dual \cap M]} 
      =&\;
      1 + 
      {\textstyle \sum_{i=1}^8} x_i + 
      \\&\;
      x_1 x_3 + 
      x_1 x_5 +
      x_2 x_6 + 
      x_3 x_7 + 
      x_3 x_5 + 
      x_6 x_8 + 
      x_1 x_3 x_7 + 
      \\&\;
      x_3 x_6 
      \big(
      x_1 + x_3 + x_4 + x_5 + x_6 + x_7 + x_8 + x_3 x_6
      \big)
      \\
      P_\infty\big|_{\Spec \C[\sigma^\dual \cap M]} 
      =&\;
      x_3 x_6
      .
    \end{split}
  \end{equation}
  A straightforward computation~\cite{GPS05} yields that $P_1 =
  P_0+P_\infty$ is transverse. Therefore, any sufficiently generic
  $P_\varphi$ is transverse.
\end{enumerate}
This proves that a generic hypersurface $X_\varphi$ is smooth. For example,
$X_1$ is smooth.

The (non-generic) hypersurfaces $X_0$ and $X_\infty$ are both singular
because the special polynomials $P_0$ and $P_\infty$ fail to be
transverse. Moreover, $X_0$ has additional orbifold singularities
because $\Xt_0$ passes through $12$ out of the $16$ $g_4^2$-fixed
points.\footnote{The $4$ points that are fixed by $g_4^2$ and missed
  by $\Xt_0$ are the $4$ fixed points of $g_4$.} Finally, $X_\infty$
has an additional singularity because $\Xt_\infty$ passes through the
singular points of the ambient toric variety. A complete description
of the complex structure moduli space will be given
elsewhere~\cite{H11mirror}.

\subsection{Hodge Numbers}

A generic (not necessarily symmetric) Calabi-Yau hypersurface in the
toric variety $\CP_\nabla$ has Hodge numbers
\begin{equation}
  h^{pq}\big(\Xt\big)
  =
  \vcenter{\xymatrix@!0@=6mm@ur{
      1 &  0 &  0 & 1 \\
      0 & 20 & 20 & 0 \\
      0 & 20 & 20 & 0 \\
      1 &  0 &  0 & 1 
    }}
  .
\end{equation}
Unsurprisingly, it is self-mirror ($h^{11}=h^{21}$) since the 24-cell
is a self-dual polytope. All complex structure deformations of $\Xt$
are represented by deformations of the defining polynomial. Therefore,
the complex structure deformations of $G$-symmetric threefolds $\Xt$
are necessarily parametrized by the $G$-invariant polynomials. As we
have seen in eq.~\eqref{eq:aKinv}, there is a one-dimensional family
$P_\varphi$ of invariant polynomials. Therefore,
\begin{equation}
  h^{21}(X) = h^{21}\big(\Xt\big)^G = 1
\end{equation}
The Euler number $0=\chi(\Xt)=\chi(X) = 2 h^{11}(X)-2h^{21}(X)$ then
implies that $h^{11}(X)=1$ as well. To summarize, the Hodge diamond of
a smooth quotient $X=\Xt/G$ Calabi-Yau threefold is
\begin{equation}
  h^{pq}\big(X\big)
  =
  \vcenter{\xymatrix@!0@=6mm@ur{
      1 &  0 &  0 & 1 \\
      0 &  1 &  1 & 0 \\
      0 &  1 &  1 & 0 \\
      1 &  0 &  0 & 1 
    }}
  .
\end{equation}

\section{Generalizations}
\label{sec:general}

\subsection{Permutation Orbifolds}
\label{sec:perm}

Let me start by explaining some of the motivation behind the group
action on the toric variety $\CP_\nabla$ defined by the $24$-cell. By
rewriting the Dolbeault cohomology and contracting with the covariant
constant $(3,0)$-form, the $h^{21}$ complex structure moduli of the
Calabi-Yau hypersurface $\Xt$ correspond to the tangent bundle-valued
cohomology group
\begin{equation}
  H^{2,1}\big(\Xt\big)
  =
  H^1\big(\Xt, \wedge^2 T^* \Xt\big)
  =
  H^1\big(\Xt, T\Xt\big)
  .
\end{equation}
Now, the tangent bundle is a subbundle of the tangent bundle of the
ambient space restricted to $\Xt$,
\begin{equation}
  0 
  \longrightarrow
  T\Xt
  \longrightarrow
  T\CP_\nabla|_\Xt
  \longrightarrow
  \Osheaf(-K_\nabla)|_\Xt
  \longrightarrow
  0
  .
\end{equation}
This leads to the long exact sequence 
\begin{equation}
  \cdots
  \longrightarrow
  H^0\big(\Xt,~\Osheaf(-K_\nabla)|_\Xt\big)
  \longrightarrow
  H^1\big(\Xt,~T\Xt\big)
  \longrightarrow
  H^1\big(\Xt,~T\CP_\nabla|_\Xt\big)
  = 0
\end{equation}
of cohomology groups, where I used the result for the tangent-bundle
valued cohomology of the toric variety from \autoref{sec:TP}. We see
that, \emph{to minimize the surviving complex structure moduli on a
  free quotient $X=\Xt/G$, one needs to minimize the number of
  $G$-invariant sections of the anticanonical bundle}. 

The invariant sections of the (pull-back of the) anticanonical bundle
are computed by the cohomology long exact sequence
\begin{equation}
  0
  \longrightarrow
  \underbrace{
    H^0\big(\CP_\nabla,~ \Osheaf \big)
  }_{\simeq \C}
  \longrightarrow
  \underbrace{
    H^0\big(\CP_\nabla,~ \Osheaf(-K_\nabla) \big)
  }_{\simeq \C^{\Delta \cap M}}
  \longrightarrow
  H^0\big(\Xt,~ \Osheaf(-K_\nabla)|_\Xt \big)
  \longrightarrow
  0
  .
\end{equation}
The simplest group actions on toric varieties are toric actions, that
is, subgroups of the maximal torus. But for these to admit a
fixed-point free action on a Calabi-Yau hypersurface is very
rare~\cite{MR2282962}. In particular, there is no such action on
$\CP_\nabla$. Given that the toric group actions do not suffice, one
is led to consider permutation actions on the homogeneous coordinates
coming from symmetries of the polytope. Note that the group elements,
represented by orthogonal matrices, act in the same way on the dual
polytopes $\nabla\in N$ and $\Delta\in M$ in order to preserve the
inner product. Since we can identify
\begin{equation}
  H^0\big(\CP_\nabla,~ \Osheaf(-K_\nabla) \big) 
  =
  \C^{\Delta \cap M}
\end{equation}
with the integral points of the polytope $\Delta$, there are always at
least two invariant sections of the anticanonical bundle: The section
corresponding to the origin of $M$, and the sum over one orbit of a
vertex of $\Delta$. If the vertices form more than one orbit, then
there are more invariant sections. Hence, one is led to search for
subgroups of $\Aut(\Delta) = \mathop{\mathrm{Weyl}}(F_4)$ that act
simply transitively,\footnote{It is, a priori, not impossible for
  groups with $|G|>24$ to act freely. The group action can then still
  be transitive but not simply transitive. Therefore, there is some
  $g\in G$ that fixes some index $i$. The corresponding divisor
  $z_i=0$ is then mapped to itself by $g$, yet not forbidden by the
  Stanley-Reisner ideal. Hence, we cannot as easily conclude that all
  fixed point sets are discrete.} that is, with a single orbit of
length $24$, on the vertices of $\nabla$. A quick
computation~\cite{GAP4} reveals that, up to conjugacy, there are $22$
different subgroups of order $24$. Of these, $4$ act simply
transitively on the vertices. The groups are $2$ subgroups isomorphic
to $SL(2,3)$ that are not conjugate to each other, a
semidirect\footnote{To fix notation, I will say that $G$ is a
  semidirect product of $N$ and $H$, written $N\rtimes H$, if there is
  a group homomorphism $G\to H$ with kernel $N$. For example,
  $SL(2,3)$ can be written as a semidirect product $SL(2,3)=Q_8
  \rtimes \Z_3$.} product $\Z_3\rtimes \Z_8$ with GAP id
\texttt{[24,1]}, and $\Z_3\times Q_8$.

Thus far, I only showed that the group action would lead to $h^{21}=1$
provided that the quotient $X=\Xt/G$ is smooth. But we are not
guaranteed that it is smooth, and one must check it
case-by-case. However, note that the invariant equations for all of
these four group actions are the same, namely the one-parameter family
$P_\varphi$ defined in eq.~\eqref{eq:Ptdef}. Hence, the covering
Calabi-Yau manifold is always the same one-parameter family
$\Xt_\varphi$. We have already shown in \autoref{sec:Smooth} that a
generic hypersurface does not meet the ambient singularities and is
transverse. Therefore, the covering space is generically smooth and
one only has to check that the group action is free on $\Xt$. That is,
the fixed point set in $\CP_\nabla$ must not meet the hypersurface for
all $g\in G-\{1\}$. Direct computation shows that the other
$SL(2,3)$-subgroup, that is the one we have not used in
\autoref{sec:SL23}, has fixed points.\footnote{In fact, the other
  $SL(2,3)$-action fixes surfaces in $\CP_\nabla$ intersecting $\Xt$
  in curves.} The remaining two groups have, like the first $SL(2,3)$,
only isolated fixed points in the maximal torus and no fixed points
outside of the maximal torus. Hence, all three groups act without
fixed points on a generic hypersurface $\Xt$.

To summarize, there are really three different free quotients of a
generic invariant hypersurface $\Xt\subset \CP_\nabla$. The three
quotients $X_i = \Xt/G_i$ are Calabi-Yau manifolds with the same Hodge
numbers $h^{11}=h^{21}=1$, but different fundamental groups. To remind
ourselves, the first one was
\begin{equation}
  G_1 = G = \langle g_3, g_4 \rangle
  \simeq SL(2,3)
  ,
\end{equation}
see eq.~\eqref{eq:g3g4perm} for the definition of the permutations
$g_3$ and $g_4$. The second group can be written as a semidirect
product
\begin{equation}
  G_2 = \langle g_3, g_8 \rangle
  \simeq
  \Z_3 \rtimes \Z_8
\end{equation}
with
\begin{equation}
  g_8 \eqdef
  (1,2,20,12,16,15,13,5)(3,10,8,11,14,7,9,6)(4,19,21,22,24,23,17,18)
  .
\end{equation}
Finally, the third group can be written as a direct product
\begin{equation}
  G_3 = \langle g_3, i, j \rangle
  \simeq
  \Z_3 \times Q_8
\end{equation}
with 
\begin{equation}
  \begin{split}
    i =&
    (1,6,16,11)(2,8,15,9)(3,5,14,12)(4,23,24,19)(7,13,10,20)(17,18,21,22)
    ,
    \\
    j =&
    (1,7,16,10)(2,3,15,14)(4,22,24,18)(5,8,12,9)(6,20,11,13)(17,23,21,19)
    .
  \end{split}
\end{equation}
The three different groups have a common subgroup
\begin{equation}
  \Z_6 = \langle g_3, g_4^2 \rangle
  = G_1 \cap G_2 \cap G_3
\end{equation}
which is normal in $G_2$ and $G_3$, but not in $G_1$.

\subsection{Subgroups and Partial Quotients}
\label{sec:sub}

The three freely acting groups $G_1$, $G_2$, and $G_3$ on $\Xt$ have a
various subgroups $H$, each of which acts freely and gives rise to another
Calabi-Yau threefold $X_H = \Xt/H$. Up to conjugation, the subgroups
are
\begin{equation}
  \begin{split}
    G_1 \supset&\;
    1,\; \Z_2,\; \Z_3,\; \Z_4,\; \Z_6,\; Q_8,\; SL(2,\;3)
    ; \\
    G_2 \supset&\;
    1,\; \Z_2,\; \Z_3,\; \Z_4,\; \Z_6,\; 
    \Z_8,\; \Z_{12},\; \Z_3 \rtimes \Z_8
    ; \\
    G_3 \supset&\;
    1,\; \Z_2,\; \Z_3,\; 
    \Z_4^{(1)},\; \Z_4^{(2)},\; \Z_4^{(3)},\; 
    \Z_6,\; Q_8,\; 
    \Z_{12}^{(1)},\; \Z_{12}^{(2)},\; \Z_{12}^{(3)},\; 
    \Z_3 \times Q_8
    .
  \end{split}
\end{equation}
Note that $G_3$ has three different $\Z_4$ and three different
$\Z_{12}$ subgroups that are not conjugate to each other. The
corresponding free quotients of $\Xt$ are not related by a symmetry of
the covering space, and might be different manifolds.

Having identified the different subgroups $H\subset G_i$, $i=1,2,3$,
one would like to know the Hodge numbers of the partial quotient
Calabi-Yau threefolds. Again because the Euler number
vanishes, it suffices to compute
\begin{equation}
  h^{21}(X_H) = 
  h^{21}\big(\Xt / H\big) = 
  \dim H^1\big(\Xt\big)^H
  .
\end{equation}
From the computation of the tangent bundle cohomology in
\autoref{sec:TP}, one can identify
\begin{equation}
  H^1\big(\Xt\big)
  =
  \C^{\fan(1)}
  -
  N \otimes_\Z \C
\end{equation}
as $H$-representations. That is,
\begin{table}
  \centering
  \begin{tabular}{cc|l}
    $h^{11}(X_H)$ & 
    $h^{21}(X_H)$ & 
    $H$ 
    \\ \hline
    $20$ & $20$ & $1$ \\
    $12$ & $12$ & $\Z_2$ \\
    $8$  &  $8$ & $\Z_3$ \\
    $6$  &  $6$ & $\Z_4$ \\
    $4$  &  $4$ & $\Z_6$ \\
    $3$  &  $3$ & $\Z_8$, $Q_8$ \\
    $2$  &  $2$ & $\Z_{12}$ \\
    $1$  &  $1$ & 
    $SL(2,3)$, 
    $\Z_3 \rtimes \Z_8$, 
    $\Z_3 \times Q_8$ 
  \end{tabular}
  \caption{Fundamental groups $\pi_1(X_H)=H$ and Hodge numbers of the
    various free quotients of the Calabi-Yau hypersurface $\Xt$ in the toric
    variety $\CP_\nabla$.}
  \label{tab:final}
\end{table}
\begin{itemize}
\item $\C^{\fan(1)}$ is the $24$-dimensional representation spanned
  by the $24$ vertices of the polytope $\nabla$, and
\item $N\otimes_\Z \C$ is the $4$-dimensional matrix representation of
  the group.
\end{itemize}
It is then an easy exercise to compute the invariant cohomology
groups. It turns out that the representation on $H^1(T\Xt)$ only
depends on the group, and not on the details of the permutation
action. The resulting Hodge numbers, for the different groups, are
listed in \autoref{tab:final}.

\appendix

\section{Cohomology of the Tangent Bundle}
\label{sec:TP}

\subsection{Of the Ambient Toric Variety}

The tangent bundle of a toric variety has a monad presentation
\begin{equation}
  0 
  \longrightarrow
  (n-d) \Osheaf
  \longrightarrow
  \bigoplus_{i=1}^n \Osheaf(V(z_i))
  \longrightarrow
  T\CP_\nabla
  \longrightarrow
  0
\end{equation}
In our particular case, one only has to be careful about the
singularities. For example, $\Osheaf(V(z_i))$ is not a line bundle,
but only a reflexive sheaf. Nevertheless, we can apply exact
sequences. Using the standard toric algorithm for the cohomology of
Weil divisors~\cite{CoxLittleSchenck}, I find
\begin{equation}
  \begin{gathered}
    h^\bullet\big(\CP_\nabla, \Osheaf\big) 
    =
    h^\bullet\big(\CP_\nabla, \Osheaf(V(z_i))\big)
    = 
    (1,\;0,\;0,\;0,\;0),
    \\ 
    h^\bullet\big(\CP_\nabla, \Osheaf(K_\nabla) \big) 
    =
    (0,\;0,\;0,\;0,\;1),
    \\ 
    h^\bullet\big(\CP_\nabla, \Osheaf(V(z_i))\otimes 
    \Osheaf(K_\nabla) \big)
    = 
    (0,\;0,\;0,\;0,\;0).    
  \end{gathered}
\end{equation}
Therefore,
\begin{equation}
  \begin{split}
    h^\bullet\big(\CP_\nabla,~ T\CP_\nabla\big)
    =&\;
    (4,\;0,\;0,\;0,\;0),
    \\
    h^\bullet\big(\CP_\nabla,~ T\CP_\nabla\otimes \Osheaf(K_\nabla) \big)
    =&\;
    (0,\;0,\;0,\;20,\;0).    
  \end{split}
\end{equation}
The restriction to a smooth anticanonical hypersurface $\Xt$ can be
computed from the short exact sequence
\begin{equation}
  \label{eq:restrXt}
  0 
  \longrightarrow
  T\CP_\nabla\otimes K_\nabla
  \longrightarrow
  T\CP_\nabla
  \longrightarrow
  T\CP_\nabla|_\Xt
  \longrightarrow
  0
  ,
\end{equation}
and I find
\begin{equation}
  h^\bullet\big(\Xt,~ T\CP_\nabla|_\Xt\big)
  =
  (4,\;0,\;20,\;0)
  .
\end{equation}

\subsection{Of the Hypersurface}

First, we need the cohomology of the restriction
$\Osheaf(-K_\nabla)|_\Xt$ of the anticanonical bundle. Analogous to
eq.~\eqref{eq:restrXt}, I find
\begin{equation}
  h^0\big(\Xt,~ \Osheaf(-K_\nabla)|_\Xt\big)
  =
  h^0\big(\CP_\nabla,~ \Osheaf(-K_\nabla)\big)
  -
  h^0\big(\CP_\nabla,~ \Osheaf\big)
  =
  25 - 1
  = 
  24
  ,
\end{equation}
and all higher cohomology groups vanish. Therefore, the tangent bundle
of the hypersurface,
\begin{equation}
  0 
  \longrightarrow
  T\Xt
  \longrightarrow
  T\CP_\nabla|_\Xt
  \longrightarrow
  \Osheaf(-K_\nabla)|_\Xt
  \longrightarrow
  0
\end{equation}
has cohomology groups
\begin{equation}
  h^\bullet\big(\Xt,~T\Xt\big) = 
  (0,\;20,\;20,\;0)
\end{equation}

\bibliographystyle{utcaps} 
\renewcommand{\refname}{Bibliography}
\addcontentsline{toc}{section}{Bibliography} 
\bibliography{Main}

\end{document}